\documentclass[iop]{emulateapj}

\usepackage{graphicx}
\usepackage{subfigure}
\usepackage{epsf}
\usepackage{longtable}
\usepackage{amsmath}

\shorttitle{}
\shortauthors{}

\newcommand{\HI}{{H{~\footnotesize I }}}

\begin{document}

\title{Discovery of a Small Central Disk of CO and \HI in the Merger
 Remnant NGC 34 }

\author{Ximena Fern\'{a}ndez\altaffilmark{1}, A.O. Petric\altaffilmark{2}, Fran\c{c}ois Schweizer\altaffilmark{3}, J.H. van Gorkom\altaffilmark{1}}
\altaffiltext{1}{Department of Astronomy, Columbia University, New York, NY 10027, USA; ximena@astro.columbia.edu}
\altaffiltext{2}{Department of Astronomy, California Institute of Technology, Pasadena, CA 91125, USA}
\altaffiltext{3}{Carnegie Observatories, 813 Santa Barbara Street, Pasadena, CA 91101, USA}

\begin{abstract}
We present CO(1-0) and  \HI(21-cm) observations of the central region of the wet merger remnant NGC 34. The Combined Array for Research in Millimeter-wave Astronomy (CARMA) observations detect a regularly rotating disk in CO with a diameter of 2.1 kpc and a total molecular hydrogen mass of ($2.1 \pm 0.2) \times10^9~M_\odot$.  The rotation curve of this gas disk rises steeply, reaching maximum velocities at $1\arcsec\ $(410 pc) from the center. Interestingly, \HI observations done with the Karl G. Jansky Very Large Array show that the absorption against the central continuum has the exact same velocity range as the CO in emission. This strongly suggests that the absorbing \HI also lies within $1\arcsec$ from the center, is mixed in and corotates with the molecular gas. A comparison of \HI absorption profiles taken at different resolutions ($5\arcsec-45\arcsec$) shows that the spectra at lower resolutions are less deep at the systemic velocity.  This provides evidence for \HI emission in the larger beams, covering the region from 1 kpc to 9 kpc from the center. The central rapidly rotating disk was likely formed either during the merger or from fall-back material. Lastly, the radio continuum flux of the central source at mm wavelengths ($5.4\pm1.8$  mJy) is significantly higher than expected from an extrapolation of the synchrotron spectrum, indicating the contribution of thermal free-free emission from the central starburst.
\end{abstract}


\section{Introduction}

Gas-rich mergers are key to understanding the hierarchical growth and
evolution of galaxies \citep{White78}.  Much progress has been made to
understand the overall role of mergers in the formation of early-type
galaxies, and it is now accepted that many are a product of gas-rich
mergers \citep[e.g.,][]{Schweizer98}. There are still some
uncertainties on how exactly the transformation takes place,
especially how a gas-rich system ultimately rids itself of gas.
These details are what possibly dictate whether the merger remnant
becomes an elliptical or whether a new disk forms, as seen in some
observations  and simulations of gas rich
mergers \citep{DS13,Barnes02}. The possible presence of a starburst or AGN further complicates
the picture.  While gas can fuel both starbursts and AGNs, these
phenomena may cause outflows and remove the gas from the inner parts.
It is thus important to study the gas in the central regions of
mergers in  its molecular and atomic phases to help predict the
ultimate fate of the system.

NGC 34 ($=$ NGC 17 $=$ Mrk 938) is an ideal candidate to study these
processes since it is gas-rich and hosts  both a starburst and an AGN.
Throughout the paper, we adopt a distance of 85.2 Mpc, derived for
$H_0 = 70$ km s$^{-1}$ Mpc$^{-1}$ by \citet[hereafter SS07]{SS07}.
The infrared luminosity of this
object is $\log (L_{IR}/L_{\odot})=11.61$ \citep{Chini92}, placing it
in mid-range of the classically defined Luminous Infrared Galaxies
(LIRGs; $11.0 < \log L_{IR}/L_{\odot} < 12.0$; \cite{Soifer87}).
In general, high IR luminosities are due to dust grains being heated by
a combination of starburst and AGN, with higher values indicating a
greater AGN contribution \citep{SandersMirabel96}. The IR luminosity
of NGC 34 suggests that the starburst is the dominant phenomenon
in the center, with a minor AGN component
 \citep{Goncalves99,Prouton04}.  This has been confirmed by a
detailed study of NGC 34's IR spectrum done by \citet{Esquej12}, where the
AGN bolometric contribution to the total infrared luminosity is
estimated to be $2_{-1}^{+2}\%$.  In addition, the IR images show
that the starburst takes place in the inner $0.5-2$ kpc of the merger
remnant.  These authors find, however, that the presence of an AGN
is needed to explain the hard X-ray luminosity.

An optical study by SS07 showed that NGC 34 is the  likely product of
an unequal-mass merger, with a rich system of young globular clusters.
Deep optical images show two long tidal tails and structures such as
ripples, jets, and shells, all typical of merger remnants
\citep{Schweizer98}. In addition to this, the surface-brightness profiles
 in $B$, $V$, and $I$ indicate the presence of a blue exponential
stellar disk with a scale length of $a=3.3$ kpc, possibly formed from
gas falling back into the
merger remnant. The optical spectrum features a blueshifted Na I D
doublet in absorption, indicative of a mean gas outflow velocity of
$-620 \pm 60$ km s$^{-1}$ and a maximum velocity of $-1050 \pm 30$ km
s$^{-1}$.

Earlier, we performed a first imaging study of the \HI and radio continuum to better
understand the fate of cold gas during the merger \citep[][hereafter
 Radio I]{Fernandez10}.  We detected $7.2 \times 10^9~M_\odot$ of \HI
gas, which is mostly distributed along the tails, and saw evidence for
gas settling onto the blue exponential disk found by SS07. The most
puzzling part of our study was the detection of a strong \HI absorption
feature of $\sim 500$ km s$^{-1}$ width at the systemic velocity, covering
both blueshifted and redshifted velocities  and seen against the
strong unresolved central continuum source of 62.4 mJy. We presented two
 possible
scenarios to explain the puzzling absorption: either we  were probing
an \HI disk in absorption against an extended continuum source so that we
 saw both the blue and redshifted part in absorption, or we  were
seeing gas associated with the tidal tails in projection against the
continuum.

In this follow-up radio study, we present new CO imaging and \HI
absorption observations obtained at higher angular resolution and
with a wider velocity coverage. We seek to determine which of the
 above two scenarios explains the puzzling absorption
better by comparing the distribution and kinematics of the CO to
the \HI absorption data. In addition, the new continuum observations at mm-wavelengths allow us to probe the spectral index at high frequencies.  Lastly, both sets of new observations have
a much wider velocity coverage, enabling us to search for an atomic or
molecular counterpart to the outflow seen optically.

The structure of the paper is as follows.  We summarize the
observations and data reduction in Section 2, present our results in
Section 3, discuss what we have learned from these observations in
Section 4, and put forth our conclusions in Section 5.

\section{Observations and Data Reduction}
\subsection{CARMA Observations}
NGC 34 was observed in the CO $J=1-0$ transition at 115.2712 GHz with
the Combined Array for Research in Millimeter-wave Astronomy (CARMA) in
May 2011.  The observations were done in two configurations:
 in D array for 4.1 hours on source and in C array for 5.8 hours.
The spacings for these two
arrays are $11-150$m and $30-350$m, respectively.  The primary flux calibrator was Uranus, and
3C454.3 was used as the passband calibrator.  Both sets of observations
used a 1 GHz bandwidth centered at 113.11 GHz, corresponding to the central frequency of the single-dish observations.  This translates to 
a heliocentric velocity of 5731 km s$^{-1}$, using the optical
definition.  This setup results in a velocity coverage of 2980 km s$^{-1}$,
starting at 4238 km s$^{-1}$ and ending at 7218 km s$^{-1}$, with
150 channels with a velocity resolution of $~20$ km s$^{-1}$.

The data were reduced with  the software package Multichannel Image Reconstruction Image Analysis and Display (MIRIAD) using standard calibration procedures for each configuration \citep{miriad}. For each band, we flagged the first and last 2 channels due to poor sensitivity at the edge of the band. The phase stability was inspected and poor data were flagged for each track.  
The C and D array calibrated visibilities were then combined and the continuum was estimated from 696 (unsmoothed) signal free channels. 
 The  deconvolution task uses a Steer CLEAN algorithm \citep{CLEAN}, and we selected natural weighting (robustness parameter set to 2) to maximize the sensitivity to broad, faint structures. 
 The final combined data cube has a
synthesized beam of $2.48\arcsec \times 2.14\arcsec$ ($\sim 1$ kpc) at FWHM and a noise of
8.4 mJy beam$^{-1}$.

\begin{deluxetable}{ccc}
\tabletypesize{\small}
\tablenum{1}
\tablecolumns{3}
\tablewidth{0pc}
\tablecaption{Observation Parameters}
\tablehead{
\colhead{}&
\colhead{CO(1--0)\tablenotemark{1}}& 
\colhead{\HI\tablenotemark{2}}
}

\startdata
Flux Calibrator &  Uranus& 3C48\\
Phase Calibrator & 3C454.3  & J2357-1125  \\
Number of Channels & 150 & 256 \\
Bandwidth  (MHz) & 1000 & 32 \\ 
Synthesized Beam &  $2.48\arcsec \times  2.14\arcsec$ & $8.31\arcsec \times 4.46 \arcsec$ \\
Field of View &$1-1.7\arcmin$ &$32\arcmin$\\
Velocity Coverage  & 3000 & 7000  \\
{(km s$^{-1}$)} & &\\
Velocity Resolution    & 20 &  27 \\
(km s$^{-1}$) & &\\
Noise & 8.4   & 0.21 \\
 (mJy beam$^{-1}$chan$^{-1}$) & & \\
Column density sensitivity\tablenotemark{3} & $1.1\times10^{20}$($H_2$) & $1.7\times10^{20}$\\
 (cm$^{-2}$ chan$^{-1}$) & &

 \enddata
\tablenotetext{1}{Observed in May 2011}
\tablenotetext{2}{Observed in April 2011}
\tablenotetext{3}{In emission}
\end{deluxetable}
\vspace{0.1in}
\begin{figure*}
\begin{center}
\includegraphics[scale=0.52]{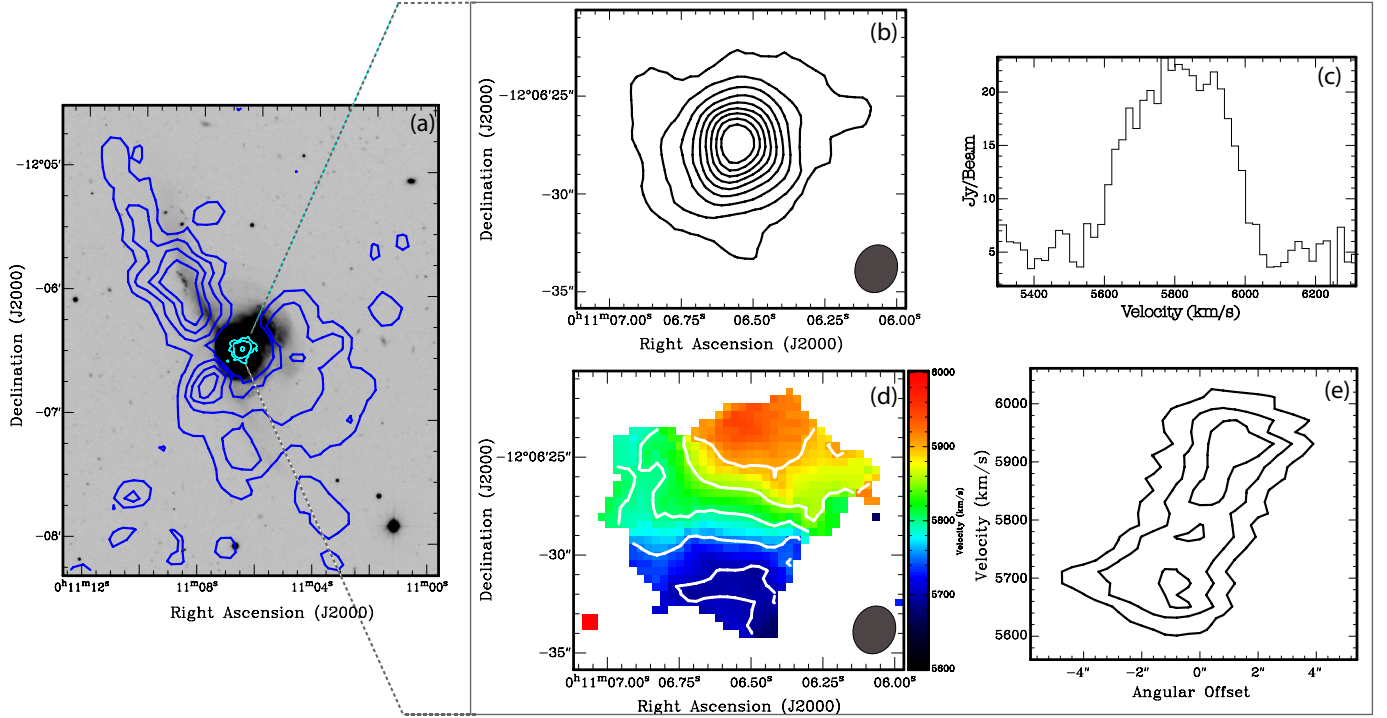}
\caption{Maps of the distribution and kinematics of the CO disk found
  in NGC 34. (a) CO emission as compared to the optical and \HI
  emission; the cyan contours show the CO emission, and the \HI
  contours (in blue; from Radio I) are drawn in levels of (8, 28, 48, 68, 108)
  $\times 10^{19}$ cm$^{-2}$ overlaid on an optical image from
  SS07. (b) CO distribution map (moment 0) contours drawn starting at 5$\%$ of the peak, in intervals of 10$\%$.  (c) The spectrum of the
  CO emission done by setting a box around the emission showing how
  much CO there is at a given velocity bin.  (d) Velocity map (moment 1)
  overlaid with isovelocity contours drawn in intervals of 50
  km s$^{-1}$. (e) Position--velocity diagram along the major axis of the
  optical disk (PA $-9^\circ$; SS07).  Note that the full range of
  velocities are not shown in these maps, we limit
  them to show a better contrast or to zoom-in in a region of interest. }
\label{fig:COdisk}
\end{center}
\end{figure*}

\subsection{VLA Observations}
NGC 34 was observed with the Karl G. Jansky Very Large Array
(VLA)\footnote{The National Radio Astronomy Observatory is a facility
  of the National Science Foundation operated under cooperative
  agreement by Associated Universities, Inc.}  in April 2011 in the B
array configuration (spacings of $0.21-11.1$km)  during two runs for
a total observing time of 7 hours. The observations used a 32 MHz bandwidth centered at the
heliocentric velocity of 5870 km s$^{-1}$, with a total of 256
channels with a resolution of $~27$ km s$^{-1}$.  This results in a
velocity coverage of 7000 km s$^{-1}$, from 2426 km s$^{-1}$ to 9421
km s$^{-1}$.

Each run was reduced with the Astronomical Imaging Processing System
(AIPS) using standard calibration procedures. The data were combined
in the \textit{uv} plane, and then  several iterations of self-calibration 
were performed to correct for  amplitude and phase errors. We first made a continuum image by
averaging 80 line-free channels, using a robustness parameter of 1 and
setting CLEAN boxes around the point source.  This image served as the
initial input model for the first run of phase self-calibration. After
each run, we made a new input model by imaging the continuum applying
the new solutions for subsequent self-calibration runs.  After three
runs of phase self-calibration and one of amplitude, we achieved a dynamic range of about 760:1 and an rms
sensitivity of 0.07 mJy beam$^{-1}$. We then applied the solutions to
the whole data set, and subtracted the continuum in the  \textit{uv}  plane by 
making a linear
fit through the line-free channels.  The \HI cube was made 
using a robustness parameter of 1, and cleaning the channels containing 
\HI absorption.
 The resulting images have an rms noise of
0.2 mJy beam$^{-1}$ and a synthesized beam of
$8.31\arcsec \times 4.46\arcsec$ ($\sim 3.5 \times 2$ kpc).

Table 1 summarizes all parameters used for the CARMA and VLA
observations.

\section{Results}
\subsection{A Central Disk of CO}
The CO observations reveal a disk of molecular gas in
the central regions of the remnant. Figure 1 is composed of various
panels showing the gas distribution and kinematics. This
CO disk lies at the center of NGC 34 and is much smaller than the
optical disk (Figure 1a). It shows a regular rotation pattern
as seen in Figures 1d and 1e. There are hints of
extended emission towards the northwest, northeast, and south
(Figure 1b).  We estimate a diameter of 2.1 kpc by measuring the
extent of the emission enclosed by the second lowest density contour,
which corresponds to 4.31 Jy km s$^{-1}$ beam$^{-1}$.

We calculate the CO luminosity with the following equation
\citep{Solomon97}:
\begin{equation}
L_{CO}=3.25\times10^7~Sdv~\nu_{obs}^{-2} D_L^2(1+z)^{-3} ,
\end{equation}
where $Sdv$ is the integrated CO flux in units of Jy km s$^{-1}$,
$\nu_{obs}$ is the observed frequency in GHz, and $D_L$ is the
luminosity distance in Mpc.  We compute the CO flux using channel maps
and get a value of $151.9~ \pm ~11.3$ Jy km s$^{-1}$, which results in
a CO luminosity of $(2.6 \pm 0.2)\times10^{9}$ K km s$^{-1}$
pc$^{-2}$.  We multiply this value by $\alpha_{CO}$ to get a molecular
hydrogen mass in  solar masses ($M_\odot$).  Here we use $\alpha=0.8$,
which is the standard conversion factor for starbursting systems
\citep{DownesSolomon98}, and get an H$_2$ mass of $2.1 \pm 0.2 \times
10^9~M_\odot$.

Single-dish CO(1--0) data \citep{Kandalyan03,Chini92,Krugel90} yield a
flux of $170 \pm 12$ Jy km s$^{-1}$, which is  12\% higher than
what we detect. This is a marginal difference, but---if significant---it
might be indicating that our observations miss an extended
component of the CO disk.  This component is hinted  at by the weak
extensions seen in Figure 1b, suggesting that observing with shorter
spacings could have detected this more extended gas.

As seen in the spectrum of Figure 1c, the CO emission spans close to  500 km
s$^{-1}$. The intensity-weighted velocity field (Figure 1d) and
the position-velocity (PV)
diagram (Figure 1e) along the major axis of the optical disk (PA
$-9^\circ$; SS07) show that the kinematics are consistent with
rotation, with the north side of the disk receding,  and the south side
approaching.  The extended emission  features mentioned
above have velocities consistent with the gas in the main body of the
disk, suggesting they are real and possibly part of a fainter component not
detected in our observations.  The PV diagram  demonstrates that the rotation
curve rises very steeply, reaching peak velocities within $1\arcsec$ from
the center, and showing a slight decrease towards the edges.
 
\begin{figure}
\begin{center}
$\begin{array}{c}
\includegraphics[trim= 2 12 5 30,clip,scale=0.4]{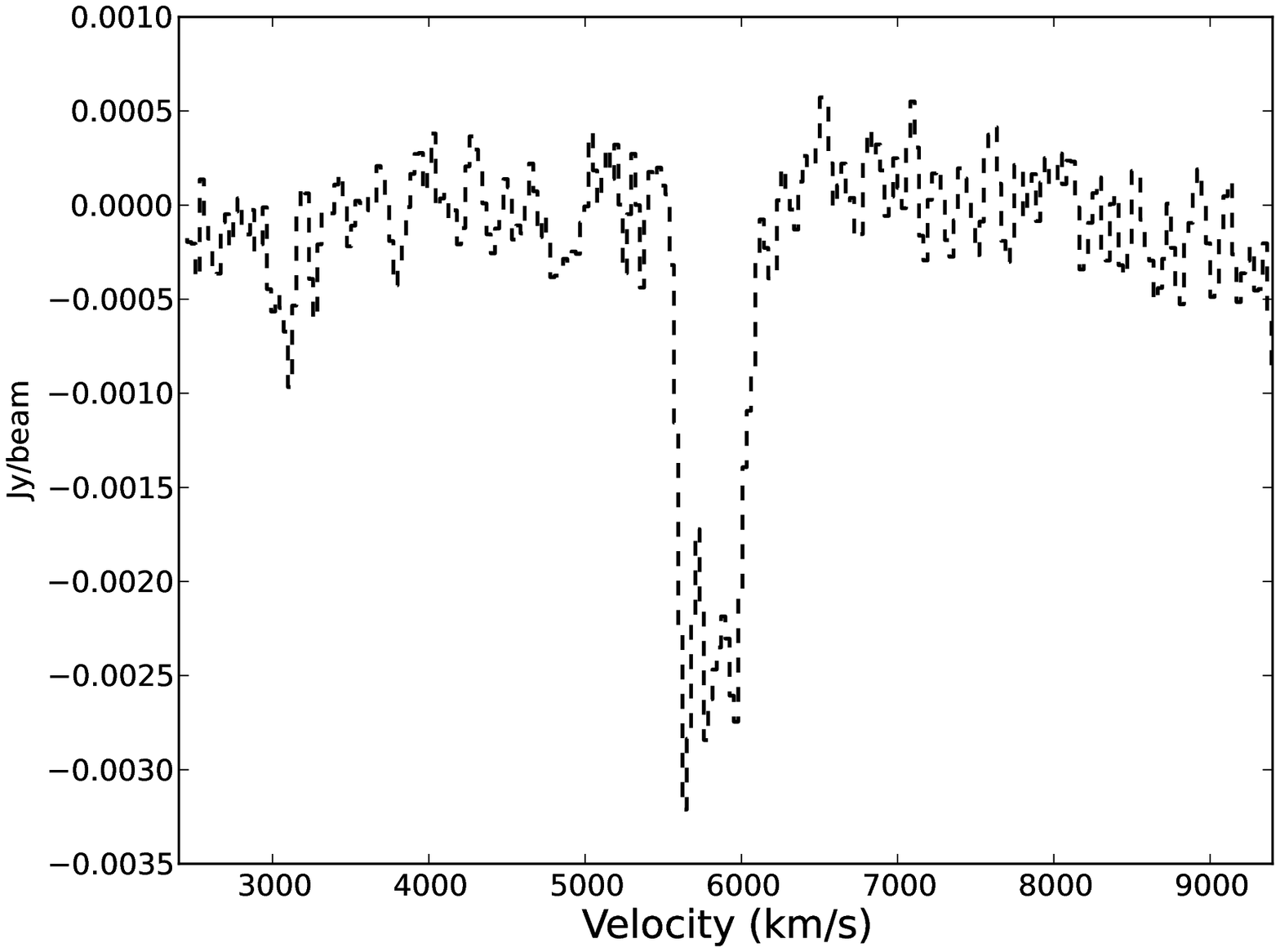}\\
\includegraphics[trim= 0 12 5 30,clip,scale=0.4]{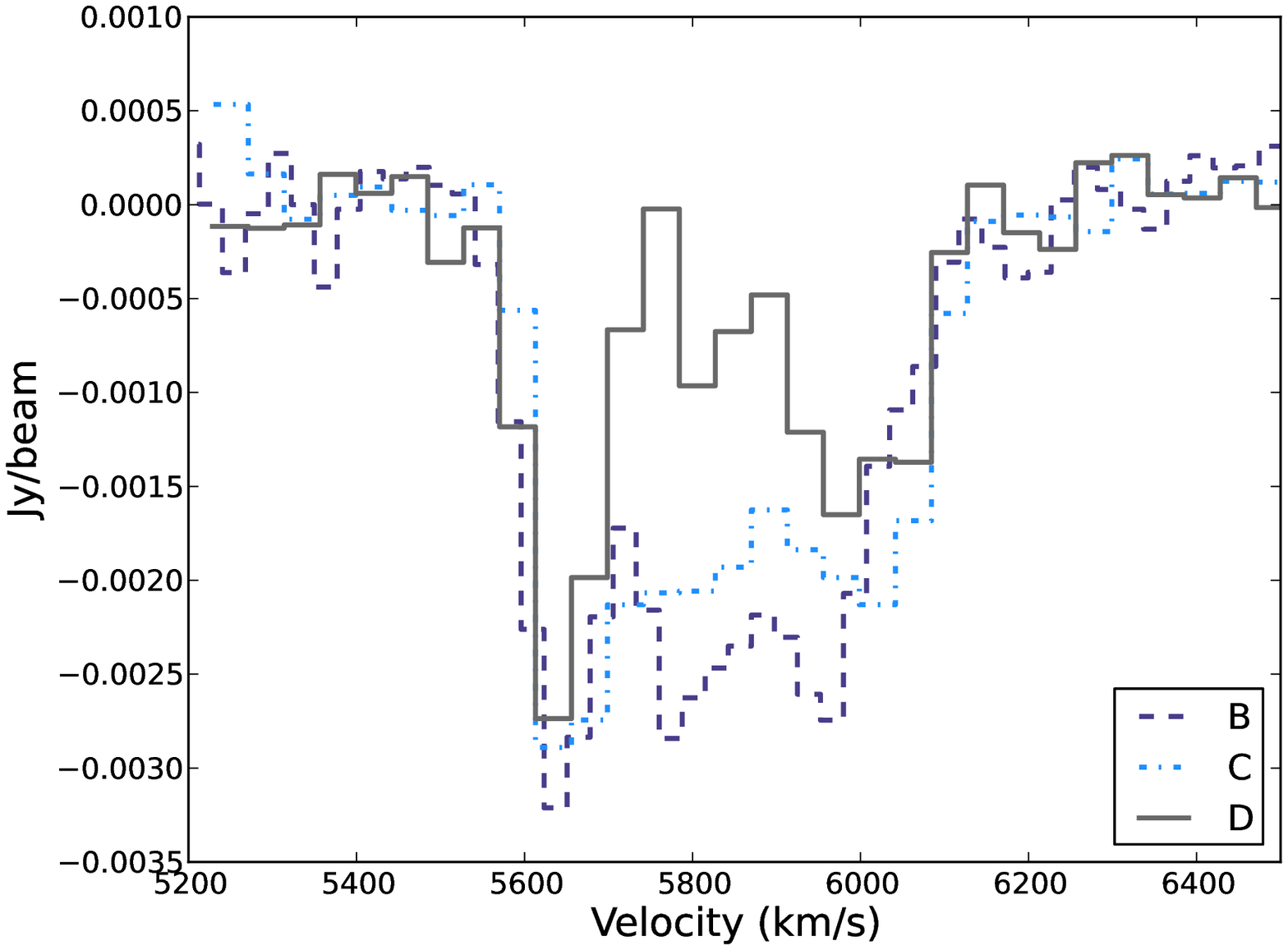}\\
\includegraphics[trim= 2 12 5 30,clip,scale=0.4]{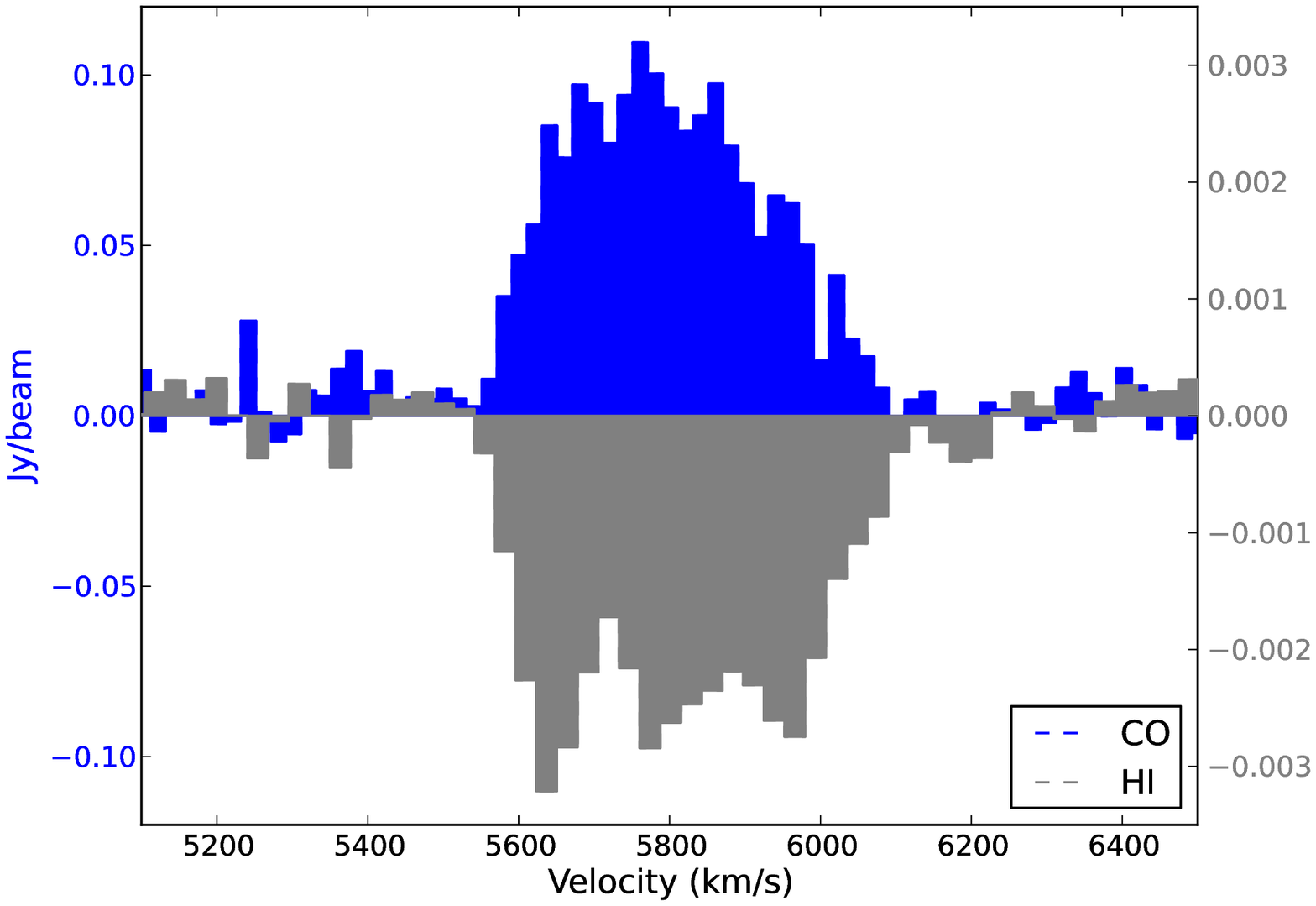}\\
\end{array}
$
\caption{\HI absorption at higher resolution and wider velocity coverage.
Top: B-array observations showing the \HI absorption at both blue shifted
and redshifted velocities.
Middle: Comparison of the absorption profile observed at three different
resolutions (DnC: 18 kpc, CnB: 6 kpc, and B: 2 kpc).
Bottom: Comparison of the \HI absorption profile with the CO emission
 spectrum.  The y-axes are different for the two, but are shown in
the same plot to compare the velocities.} 
\label{fig:HIabs}
\end{center}
\end{figure}

\subsection{High Resolution \HI Absorption}
 Our new VLA B-array observations probe the previously known \HI
absorption feature (Radio I) with a wider velocity coverage and at
higher resolution. Figure 2 consists of three panels showing these
observations, including a comparison to the absorption profiles
presented in Radio I, and to the CO observations in emission presented in
Section 3.1. The top panel  represents the \HI absorption feature seen
against the peak of the continuum emission and showing the full velocity
coverage of 2426--9421 km s$^{-1}$, with no hint of absorption seen
at the outflow velocity ($\sim$5000 km s$^{-1}$) detected by SS07 (see
next section).

The second panel compares these higher resolution (2 kpc) data
to the previous \HI observations in the DnC (18 kpc) and CnB (6 kpc)
configurations. Note that the depth of absorption is similar in the
three configurations at the extreme velocities
around 5600 km s$^{-1}$ and 6000 km s$^{-1}$, but is very
different near the systemic velocity of 5870 km s$^{-1}$.
A deep feature is seen in B array at this velocity,
which is shallower in C and nearly absent at some velocities in D. 
We  can understand this as follows:
Since the continuum source is unresolved  in B array, we see 
a combination of absorption against the continuum source plus emission 
further out within the central beam in C and D array. We can estimate how
much emission there is on different scales by looking at the difference
in the absorption profiles.  
We do this by summing the difference between the two profiles at different
resolutions for each velocity bin to get a flux.
As shown in Radio I,
there is $7.3 \times 10^8~M_\odot$ of \HI in emission between the inner
6 and 18 kpc of the remnant.  Now we can determine how much \HI in
emission there is between the inner 2 and 6 kpc by comparing the CnB
and the B array data.  We calculate an \HI mass of $2.5 \times
10^8~M_\odot$ in the velocity interval $5700-6000$ km s$^{-1}$. Since
the two data sets have different velocity resolutions, we created
velocity bins of 50 km s$^{-1}$ and then averaged the values that were in
a given bin.
 
The bottom panel compares the range of velocities seen in the \HI
absorption and the CO emission.  The plot shows that the velocity
ranges for the emission and the absorption are almost
identical. The range covered for the \HI absorption profile is  $466 \pm 27$ km s$^{-1}$  and for the CO profile is $460 \pm 20$ km s$^{-1}$, as measured at the profile base. This remarkable agreement in velocity range makes it 
plausible that the \HI and CO emissions arise from the same central
 gas disk, imaged in CO at  $2.48\arcsec \times 2.14\arcsec$ resolution (Figs 1b and 1d).   In addition, it implies that the \HI is intermixed with the molecular gas and that it also reaches maximum velocities within $\sim1\arcsec$ ($0.4$ kpc) from the center.
 
\subsection{Molecular and Atomic Gas Outflow}
We can use the current observations to search for a counterpart in \HI
and/or  CO(1--0) to the outflow detected by SS07 in the NaD I doublet.
We do not detect the outflow in either line, but can place upper limits.
We calculate the noise near the central regions of NGC 34 and around
$\sim$5000 km s$^{-1}$, consistent with the velocities of the optical 
outflow.  We use the following equation to get a mass outflow rate
upper limit \citep{Heckman00,Rupke02} for both \HI absorption and
emission:
\begin{multline}
 \dot{M}(H)=21\left(\frac{\Omega}{4\pi}\right)C_f\left(\frac{r_*}{1 ~ \rm kpc}\right)\left(\frac{N(H)}{10^{21}~ \rm cm^{-2}}\right)\\\times\left(\frac{\Delta v}{200 \rm~ km~ s^{-1}}\right)
  M_{\odot}~\rm yr^{-1} ,
\end{multline}
where $\Omega$ is the solid angle subtended by the outflow, $C_f$ is
the line-of-sight covering fraction, $r_*$ is the size of the region from
where the outflow is produced, $N(H)$ is the column density, and
$\Delta v$ is the velocity of the outflow.  For the first three quantities,
we assume $\Omega=4\pi$, $C_f=0.5$, and $r_*$= 1 kpc, which are values
consistent with outflows seen in LIRGs \citep{Rupke02}.  For the
absorption, we calculate the \HI column density assuming 5$\sigma$ and
get $4.2\times10^{20}$ cm$^{-2}$ (assuming $T_s=100$ K). Lastly, we
take $\Delta v=620$ km s$^{-1}$, which is the mean velocity of the
outflow detected by SS07.  The resulting upper limit for the mass
outflow rate is 14 $M_{\odot}~\rm yr^{-1}$ in absorption. Using the
same values, but now calculating N(\HI) in emission, the upper limit
is about 130 $M_{\odot}~\rm yr^{-1}$. This large number is a consequence
of the poor surface brightness sensitivity in B array. 

>From the CO emission, we can calculate an upper limit for the outflow
rate by dividing the molecular mass by the dynamical timescale
\citep[e.g.,][]{Alatalo11}. We compute an upper limit for the
molecular-gas mass using 5$\sigma$ and the velocity of the optical
outflow, resulting in $6.2 \times 10^7~M_\odot$.  We can estimate the
dynamical time by assuming a size of 1 kpc for the wind emitting
region and the optical velocity, which results in a timescale of
only 1.6 Myr.  These two values yield an upper limit for the CO mass outflow
rate of  about 40 $M_{\odot}~\rm yr^{-1}$.

\begin{figure}
\begin{center}
\includegraphics[trim= 25 5 35 35,clip,scale=0.49]{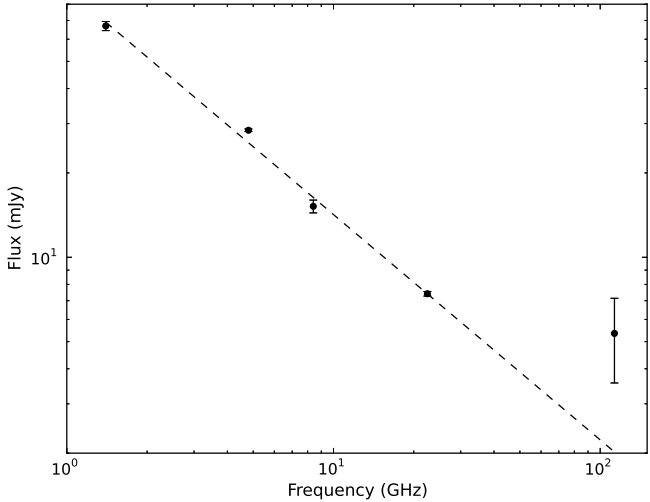}
\caption{Flux measurements at different frequencies, with the first four
measurements taken from \citet{Clemens08} and the measurement at
113 GHz showing the new observations presented here.  The line
 marks the least-squares fit to the first four measurements (slope of $-0.8$), but is
extended to demonstrate that the spectral index changes at higher
frequencies.}
\end{center}
\end{figure}

\subsection{Central continuum}
This study includes new continuum images of the central components of
the merger remnant.  The VLA B-array observations show a slightly resolved
point source with a flux of $60.6 \pm 2.0$ mJy, consistent with the measurement presented in Radio I.

The CARMA observations reveal a faint central continuum source at millimeter wavelengths with a flux of  $5.4 \pm 1.8$ mJy. This measurement allows us to probe the spectral index at high frequencies.  At low frequencies ($1-10$ GHz), the spectrum  for star-forming galaxies is usually dominated by synchrotron radiation, with typical values of $\alpha=-0.8$.  At higher frequencies ($10-200$ GHz), free-free emission starts to dominate and the spectral index is $\alpha=-0.1$. Dust emission takes over at frequencies above 200 MHz with $\alpha=1.5$  \citep{Condon92}.

Figure 3 shows the radio spectrum for NGC 34 between 1.4 GHz and 113.11 GHz. The first four values in the range of $1.4 - 22.5$ GHz were reported by  \citet{Clemens08}, and the one at 113.11 GHz is the measurement presented here.   
We fit a line through the first four measurements and extend it to 113.11 GHz to better show that the spectrum flattens at higher frequencies. The fit of the line yields a slope of $-0.8$, indicating that the emission at these frequencies is due to synchrotron radiation. The  spectral index between 22.5 GHz and 113.11 GHz is
$\alpha=-0.20\pm0.07$. The flattening of the spectrum at higher frequencies shows that synchrotron emission is no longer dominant, and free-free emission from H{\footnotesize{II}} regions takes over.

\section{Discussion}
\subsection{The Gaseous Disk in the Inner Regions of the Remnant}
 Our CO observations confirm the presence of  a 
disk in the inner regions of the remnant, which was one of the two
scenarios suggested in Radio I to explain the \HI absorption observations.
The match in  velocities for both sets of observations  (Fig.\ 2)
strongly suggests that the central \HI absorption is due to the presence of a rotating disk.
As shown in the PV diagram  of Figure 1e, the maximum
velocities are reached within only 1$\arcsec$ (0.4 kpc) from the center.  The close
match in velocity between the molecular gas and \HI absorption argues
that the absorbing gas also reaches maximum velocities within 0.4 kpc  from the
center. This is consistent with higher-resolution continuum images by \citet{Condon91} that resolve the central source and show it to be a dominant central component with an extension to the south, with a total size of ~2.5$\arcsec$ (1 kpc). This extended continuum makes it possible to trace out the \HI absorption to its maximum velocity at $1\arcsec$ from the center.

The question now is whether any of the continuum emission  could be due to the AGN. 
Several recent studies have calculated the AGN contribution to be small ($1-10\%$) in NGC 34 \citep[e.g.,][]{Vega08,Esquej12, Murphy13}. In addition to this, VLBI observations report a single-baseline detection \citep{Lonsdale93}, which suggests that this merger remnant does not host a strong point source due to the AGN, and its continuum is likely dominated by the starburst.

\subsection{The Fate of Gas in the Merger}
The optical study by SS07 and the radio observations presented here and in Radio I have shown different components of the new disk formed after the merger.  SS07 find a blue stellar disk, with spiral
structure extending out to a radius of $\sim$3.3 kpc, that seems to have formed $\sim$400 Myr ago and is embedded in a
red spheroid.  In Radio I, we found that the
\HI from the northern tidal tail is falling back and continuing to feed
the outer regions of this optical disk, favoring the idea of inside-out
growth.  On smaller scales, we have shown in this study that there is
a central disk of CO and \HI of  about 2.1 kpc in diameter. 
In addition, we find evidence for more extended \HI emission further out as shown by the comparison of the absorption profiles at different resolutions. 

Observations and simulations of mergers suggest that about 50\% of the gas
gets quickly moved to the central regions, losing its angular momentum
due to cloud-cloud collisions,  while the other 50\% gets moved to large distances
from the center, but remains bound and will eventually fall back \citep{Hibbard95,Hibbard96,Barnes02}. In NGC 
34 we see evidence for both  processes, and at the present time there are comparable amounts of cold gas in the tidal tails and the center.  We know the blue stellar disk formed $\sim$400 Myr ago
from gas settling toward the center during the merger.  The central gaseous disk presented here could have formed during the merger or more recently from fall-back material.  The major axes of the gaseous disk and the blue stellar disk have almost identical position angles, indicating that the two disks either have the same formation mechanism or are dynamically linked.  The current SFR in the inner 2 kpc calculated by \citet{Esquej12} from the 24$\micron$ luminosity is 42 M$_\odot$ yr$^{-1}$, which implies a consumption timescale of $\sim50$ Myr for the molecular hydrogen presented here.  This indicates that the gas in the central disk must be continuously replenished to sustain the current star formation rate.  

Lastly, both sets of observations have enough velocity coverage to permit searching for
a possible radio counterpart to the outflow seen optically in Na D I absorption by SS07.
 However, we do not find any evidence for such a counterpart in
the atomic- or molecular-gas phase.  We are able to place approximate upper
limits on the HI mass outflow rate in both emission and absorption,
and for the CO in emission. 

\section{Summary and Conclusions}
We have presented new high-resolution HI and CO observations to
study the central regions of the merger remnant NGC 34.
Our main findings are as follows:
\begin{itemize}
\item We detect a rotating CO disk of 2.1 kpc in diameter with a
  molecular-hydrogen mass of $(2.1\pm 0.2)\times 10^9~M_\odot$.
\item The velocity width of the CO(1--0) emitting gas matches the \HI
  absorption width, indicating that the broad \HI absorption is due to
  the central gas disk and is co-spatial with the CO. 
\item The new B-array observations of the \HI absorption feature allow
  us to calculate that there is $2.5 \times 10^8~M_\odot$ of \HI in the
   annular region of $1-3$ kpc from the center.  
\item We do not detect a molecular- or atomic-gas outflow, but place
  upper limits on the mass outflow rates  for \HI in emission and
  absorption, and  for CO in emission.
\item We have obtained new continuum images of the central source at
  1.4 GHz and 113 GHz.  We find that the spectrum flattens at the highest 
  frequencies to a spectral index of $\alpha=-0.20\pm0.07$.  This
  indicates that the radio continuum at those frequencies is due to
  free--free emission from the central starburst.
\end{itemize}

\acknowledgments

Support for CARMA construction was derived from the Gordon and Betty
Moore Foundation, the Kenneth T. and Eileen L. Norris Foundation, the
James S. McDonnell Foundation, the Associates of the California
Institute of Technology, the University of Chicago, the states of
California, Illinois, and Maryland, and the National Science
Foundation. Ongoing CARMA development and operations are supported by
the National Science Foundation under a cooperative agreement, and by
the CARMA partner universities.
This work was partially supported by NSF grant  No. 1009476 to Columbia University.

\end{document}